\documentclass[twocolumn,preprintnumbers,superscriptaddress,nofootinbib,aps,prd,floatfix]{revtex4}

\usepackage{enumerate}
\usepackage{amsmath,amssymb}
\usepackage{graphicx}
\usepackage{physics}
\usepackage{bbm,slashed}
\usepackage{bbold}
\usepackage{xspace,slashed}
\usepackage{hyperref}
\hypersetup{colorlinks=true, citecolor=blue, urlcolor=blue, linkcolor=blue}
\usepackage[normalem]{ulem}
\usepackage{subfigure}

\usepackage{cancel} 
\usepackage[normalem]{ulem} 

\usepackage{xcolor}
\colorlet{MAGENTA}{magenta}
\colorlet{BLACK}{black} 

\usepackage{tabularx,booktabs} 
\newcolumntype{C}{>{$}c<{$}}
\AtBeginDocument{%
  \heavyrulewidth=.08em
  \lightrulewidth=.05em
  \cmidrulewidth=.03em
  \belowrulesep=.65ex
  \belowbottomsep=0pt
  \aboverulesep=.4ex
  \abovetopsep=0pt
  \cmidrulesep=\doublerulesep
  \cmidrulekern=.5em
  \defaultaddspace=.5em
}

\newcommand\refeq[1]{Eq.~(\ref{#1})}

\newcommand\refse[1]{Sect.~\ref{#1}}

\newcommand\citere[1]{Ref.~\cite{#1}}
\newcommand\citeres[1]{Refs.~\cite{#1}}

\newcommand{\BR}{\text{BR}}

\newcommand{\re}{\text{Re}}

\newcommand{\gev}{\text{GeV}}

\begin{document}

\providecommand{\abs}[1]{\lvert#1\rvert}
\preprint{DESY-23-203}

\title{Constraints on the trilinear and quartic Higgs couplings from
triple Higgs production\\[.2em]
at the LHC and beyond}
\begin{abstract}
	Experimental information on the trilinear Higgs boson self-coupling $\kappa_3$ and the quartic self-coupling $\kappa_4$ will be crucial for gaining insight into the shape of the Higgs potential and the nature of the electroweak phase transition. While Higgs pair production processes provide access to $\kappa_3$, triple Higgs production processes, despite their small cross sections, will provide valuable complementary information on $\kappa_3$ and first experimental constraints on $\kappa_4$. 
	We investigate triple Higgs boson production at the HL-LHC, employing efficient Graph Neural Network methodologies to maximise the statistical yield. We show that it will be possible to establish bounds on the variation of both couplings from the HL-LHC analyses that significantly go beyond the constraints from perturbative unitarity. We also discuss the prospects for the analysis of triple Higgs production at future high-energy lepton colliders operating at the TeV scale.
\end{abstract}

\author{Panagiotis Stylianou}\email{panagiotis.stylianou@desy.de} 
\affiliation{Deutsches Elektronen-Synchrotron DESY, Notkestr.~85,  22607  Hamburg,  Germany\\[0.1cm]}
\author{Georg Weiglein} \email{georg.weiglein@desy.de}
\affiliation{Deutsches Elektronen-Synchrotron DESY, Notkestr.~85,  22607  Hamburg,  Germany\\[0.1cm]}
\affiliation{II.\  Institut f\"ur  Theoretische  Physik, Universit\"at  Hamburg, Luruper Chaussee 149, 22761 Hamburg, Germany\\[0.1cm]}

\maketitle

\section{Introduction}
\label{sec:intro}

Since the discovery of a
Higgs boson with a mass of about $125\, \gev$ in 2012~\cite{ATLAS:2012yve,CMS:2012qbp}, a tremendous and ongoing effort has been enacted in order to 
gain insights into the properties and interactions 
of the detected state. Its couplings with 
third generation
fermions and weak gauge bosons, as well as the loop-induced couplings with gluons and photons, have been 
investigated in detail, indicating agreement with the predictions of the Standard Model (SM) within the present experimental and theoretical uncertainties. 
In view of the plethora of possible connections of the 
detected Higgs boson to sectors 
of physics beyond the SM (BSM), 
probing the Higgs interactions 
with respect to possible effects of BSM physics will be
of central importance at the present and future runs at the LHC and at any 
future collider.

In this context the Higgs boson self-couplings are of particular relevance, while experimentally these couplings are very difficult to access. 
Experimental information about the trilinear and quartic Higgs couplings is needed to gain insights about the shape of the Higgs potential, which will have implications for a better understanding of the electroweak phase transition in the early universe and may be instrumental for explaining the observed asymmetry between matter and anti-matter in the universe.
In the SM the Higgs potential
is given by
\begin{equation}
\label{eq:smhiggs}
	V(\Phi) = \lambda (\Phi^\dagger \Phi)^2 - \mu^2 \Phi^\dagger \Phi
\end{equation}
in terms of the single Higgs doublet field $\Phi$. 
In extended scalar sectors the potential can have a much richer structure.
While the cubic and quartic Higgs couplings arising from \refeq{eq:smhiggs} are correlated in the SM 
and can be predicted in terms of the known experimental values of the mass of the detected Higgs boson and the vacuum expectation value, large deviations from the SM predictions for the Higgs self-couplings are possible even in scenarios where the other couplings of the Higgs boson at $125\, \gev$ are very close to the SM predictions (see e.g.\ \citere{Bahl:2022jnx} for a recent discussion of this point for the case of the trilinear Higgs coupling).
Experimental constraints on the trilinear and quartic Higgs self-couplings can be expressed in terms of the 
so-called $\kappa$-framework, where $\kappa_3$ ($\kappa_4$) denotes the coupling modifier of the cubic (quartic) coupling from its SM value at lowest order, i.e.~$\kappa_i = g_i / g_i^\text{SM}$, where $g_i$ denotes the value of the coupling and $g_i^\text{SM}$ its lowest-order SM prediction, 
and $i = 3, 4$. 

The most direct probe of the trilinear Higgs coupling at the LHC is the production of a pair of Higgs bosons, where $\kappa_3$ enters at leading order (LO). Both the ATLAS~\cite{ATLAS:2022jtk} and CMS~\cite{CMS:2022dwd} collaborations determine the limits on $\kappa_3$ from both gluon fusion and weak boson fusion (WBF) from different decay channels of the Higgs boson. At next-to-leading order (NLO), the trilinear Higgs coupling contributes to the Higgs-boson self-energy and also enters in additional one-loop and two-loop diagrams in WBF and gluon fusion, respectively, enabling the possibility of an indirect measurement through single-Higgs production~\cite{Degrassi:2016wml,Maltoni:2017ims,DiVita:2017eyz,Gorbahn:2016uoy,Bizon:2016wgr}. The inclusion of single-Higgs information by the ATLAS collaboration results in the most stringent bound to date on $\kappa_3$: $\left[ -0.4, 6.3 \right]$. Triple-Higgs production is known to suffer from very small cross sections, but yields additional information on $\kappa_3$ which could be used in combination with the aforementioned searches. Furthermore, it 
can provide the first experimental constraints on the quartic Higgs coupling $\kappa_4$.

The paper is structured as follows. In~\refse{sec:unitarity} we discuss the allowed values of $\kappa_3$ and $\kappa_4$ from the perspective of perturbative unitarity and show that sizeable contributions to $\kappa_4$ can occur, especially if $\kappa_3$ deviates from the SM value. We explore in~\refse{sec:hllhc} how well the HL-LHC will be able to constrain $\kappa_3$ and $\kappa_4$ from the $6b$ and $4b2\tau$ channels. Lepton colliders are additionally explored in~\refse{sec:llcol} before conclusions are presented in~\refse{sec:conc}.

\section{Current bounds, unitarity and theoretical motivation}\label{sec:unitarity}

Besides the experimental constraints from Higgs pair and triple Higgs production processes, which will be discussed below, theoretical bounds can be placed on the Higgs self-couplings from the requirement of perturbative unitarity. In our analysis we employ the unitarity constraints obtained at tree level.

A general matrix element for $2\rightarrow2$ scattering with initial and final states $\ket{i}$ and $\ket{f}$, respectively, can be decomposed in terms of partial waves through the Jacob-Wick expansion~\cite{Jacob:1959at}
\begin{equation}
	\label{eq:jacobwick}
	{\cal{M}}_{if} = 16 \pi \sum_J (2 J + 1) a_{fi}^J ({\cal{D}}_{\lambda_i, \lambda_f}^J (\theta, \phi))^* \;,
\end{equation}
where $J$ indicates the total angular momentum of the 
corresponding amplitude,
and the $\lambda_{i,f}$ 
denote the helicities of the initial and final states. The most relevant channel at tree level for constraining $\kappa_3$ and $\kappa_4$ is 
$HH \rightarrow HH$ scattering, where the Wigner-D functions ${\cal{D}}_{\lambda_i, \lambda_f}^J$ reduce to unity for the zeroth partial wave. Conservation of probability 
leads to the requirement
that the perturbative expansion must satisfy the optical theorem, which 
can be used to obtain an
upper bound on the zeroth partial wave of
\begin{equation}
	\abs{\re\left(a_{ii}^0\right)} \leq \frac{1}{2}\;,
\end{equation}
which if violated indicates inconsistencies in the perturbative calculation. 

\begin{figure}[!h]
	\includegraphics[width=0.45\textwidth]{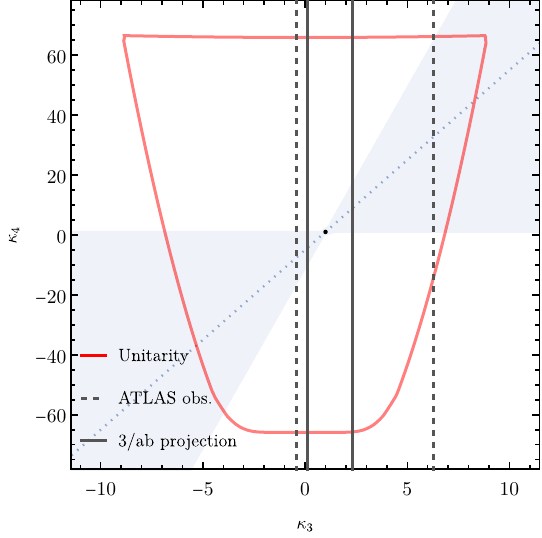}
	\caption{Bounds from perturbative unitarity on $\kappa_3$ and $\kappa_4$
 as obtained from $HH \rightarrow HH$ scattering. In addition, the current experimental bounds 
 on $\kappa_3$ are shown (black dashed lines), as well as
 the expected projections from the HL-LHC (black solid lines).
 The shaded light blue region 
 indicates where the dimension-eight contributions to $\kappa_4$ are smaller than the dimension-six ones, while the dotted blue line 
 corresponds to
 the case 
 where the dimension-eight contributions vanish,
    $\kappa_4 - 1 \simeq 6 (\kappa_3 - 1)$. \label{fig:unitarity}}
\end{figure}

The zeroth partial wave at tree level can be calculated as\footnote{We project out the zeroth partial waves from the matrix element computed through {\sc{FeynArts}}~\cite{Kublbeck:1990xc,Hahn:2000kx} and {\sc{FormCalc}}~\cite{Hahn:2001rv}, 
which is in agreement with the result of Ref.~\cite{Liu:2018peg} (see also \citeres{DiLuzio:2017tfn,Chang:2019vez,Falkowski:2019tft})
} 
\begin{equation}
	\label{eq:zerothpw}
	\begin{split}
		a_{ii}^0 =  \frac{ 3 M_H^2 \sqrt{s^2 - 4 M_H^2 s}}{ 32 \pi s (s - M_H^2) v^2} \left[  -\kappa_4 (s - M_H^2) - 3 \kappa_3^2 M_H^2\right.  \\  \left. + \frac{6 \kappa_3^2 M_H^2 (s - M_H^2)}{s - 4 M_H^2} \log\left( \frac{s}{M_H^2} - 3\right)\right] .
	\end{split}
\end{equation}
In the limit where the centre of mass energy is high,
$a_{ii}^0$ solely depends
on $\kappa_4$, while at lower energies a sizeable 
contribution from $\kappa_3$ can 
yield a peak in $a_{ii}^0$ that surpasses the allowed limit. We 
have calculated the zeroth partial wave 
for different values of $\kappa_3$ and $\kappa_4$ for a large range of energies in order to identify the 
parameter regions that are allowed by tree-level perturbative unitarity. Fig.~\ref{fig:unitarity} shows the bounds from perturbative unitarity along with the current experimental bounds on the trilinear coupling 
from ATLAS, $\kappa_3 \in [-0.4, 6.3]$ at the 95\% C.L.~\cite{ATLAS:2022jtk}, and the $95$\% combined ATLAS and CMS HL-LHC projection under the SM hypothesis, $\kappa_3 \in [0.1, 2.3]$~\cite{Cepeda:2019klc}. 

The unitarity bounds on $\kappa_4$ are significantly 
weaker than the ones on
$\kappa_3$. This 
feature can be understood
from the Effective Field Theory (EFT) perspective, 
where effects from higher-dimensional operators to the potential are incorporated
as an expansion 
in terms of inverse powers of
a UV-scale $\Lambda$~\cite{Boudjema:1995cb} (see also the discussion in~\citeres{Maltoni:2018ttu,Falkowski:2019tft})
\begin{equation}
	\label{eq:highpot}
		V_\text{BSM} = \frac{C_6}{\Lambda^2} \left( \Phi^\dagger \Phi - \frac{v^2}{2} \right)^3 + \frac{C_8}{\Lambda^4} \left(\Phi^\dagger \Phi - \frac{v^2}{2} \right)^4 + {\cal{O}}({\frac{1}{\Lambda^6})}\;.
\end{equation}
We use the convention where in the unitary gauge $\Phi = \left(0, (v + H)/\sqrt{2}\right)$, where $v$ denotes the electroweak vacuum expectation value (VEV), and $H$ is the $125$~GeV Higgs boson.
The benefit of this parameterisation of the higher-dimensional operators is that $\kappa_3$ receives corrections purely from dimension-six operators, while $\kappa_4$ only from dimension-six and dimension-eight operators (interaction vertices with more Higgs legs would additionally receive corrections from ${\cal{O}}(1/\Lambda^6)$ terms and so on). With the definitions of $\kappa_i$ as before, the coupling modifiers receive corrections
\begin{equation}
	\label{eq:k3k4rel}
    \begin{split}
        (\kappa_3 - 1) &= \frac{C_6 v^2}{\lambda \Lambda^2}\;,\\
        (\kappa_4 - 1) &= \frac{6 C_6 v^2}{\lambda \Lambda^2} + \frac{4 C_8 v^4}{\lambda \Lambda^4}\\
                       &\simeq 6 (\kappa_3 - 1) + {\cal{O}}\left(\frac{1}{\Lambda^4}\right)\;.
    \end{split}
\end{equation}
Thus, if a small correction is induced in $\kappa_3$, one should expect that in an EFT theory with high scale cutoff where the dimension-eight terms are negligible, the deviation in $\kappa_4$ from the SM expectation would be six times larger. Even if the higher-dimensional contributions are relevant, $\lvert (\kappa_4 - 1) - 6 (\kappa_3 - 1) \rvert < 6 \lvert \kappa_3 - 1\rvert$ needs to be satisfied in order to maintain a well-behaved expansion in powers of $\Lambda$. Although in this work we choose to work in all generality without any EFT assumptions on the $\kappa_3$ and $\kappa_4$ modifiers, we 
indicate the region where this condition is fulfilled in Fig.~\ref{fig:unitarity}.

\begin{figure*}[t!]
\subfigure{\includegraphics[width=8.3cm]{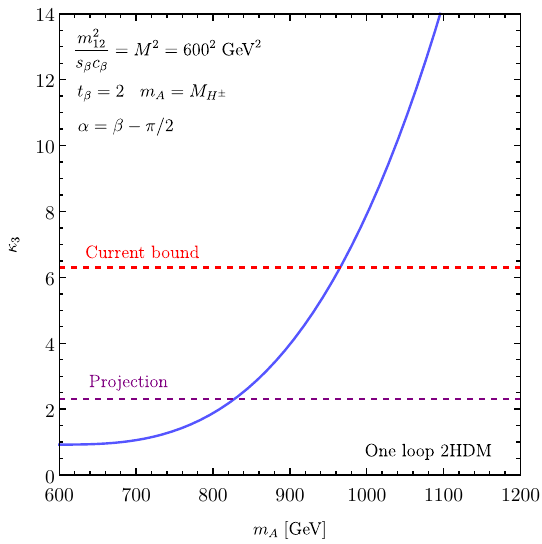}}
\hfill
\subfigure{\includegraphics[width=8.3cm]{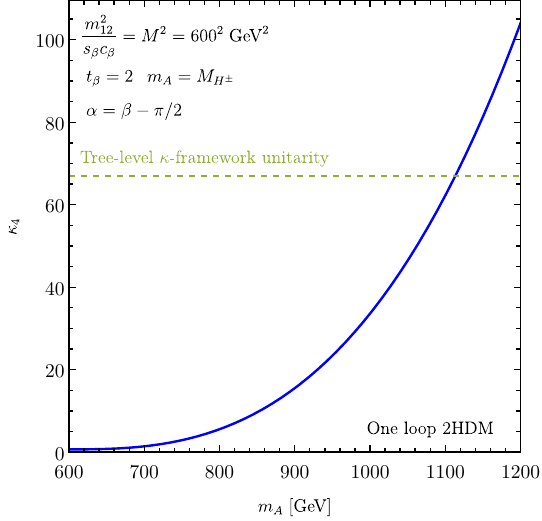}}
	\caption{The left plot shows the impact of an increasing splitting between 
 the masses of the BSM Higgs bosons on the one-loop prediction for
 the trilinear Higgs coupling $\kappa_3$,
where $M = m_H = 600$~GeV and
 $m_A = M_{H^\pm}$ 
 is varied,
 in agreement with the results of Ref.~\cite{Bahl:2022jnx} for the quoted benchmark point. On the right, the respective plot for $\kappa_4$ is shown. The shorthand notations $s_\beta$, $c_\beta$ and $t_\beta$ denote $\sin\beta, \cos\beta$ and $\tan\beta$, respectively. 
 \label{fig:loop}}
\end{figure*}

\begin{figure*}[t!]
	\includegraphics[width=8.3cm]{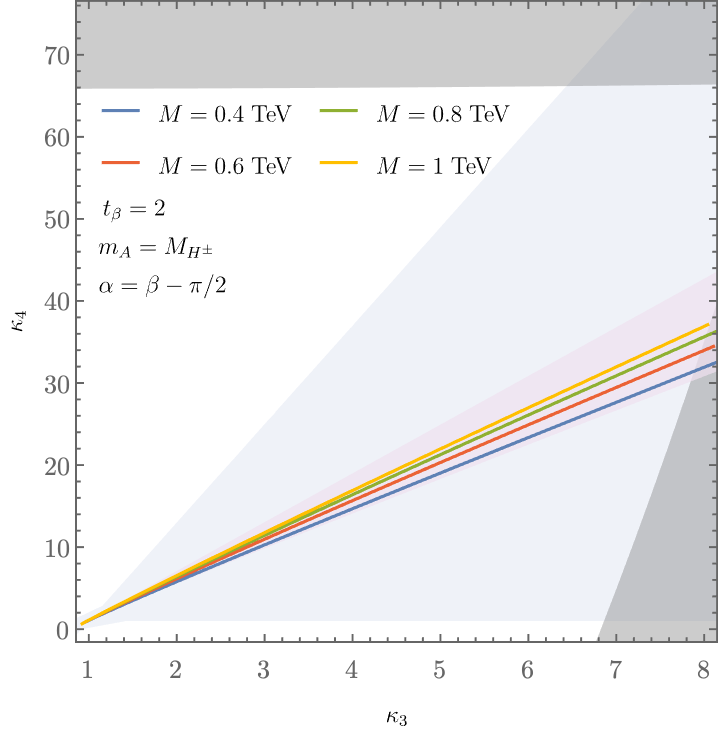}
	\caption{
 Correlation of $\kappa_3$ and $\kappa_4$ for different values of $M = m_H$ and $m_A = M_{H^\pm}$. The light purple area 
 indicates the range of 2HDM one-loop predictions for the correlation between $\kappa_3$ and $\kappa_4$ for
 $m_A, M \in [0.3, 10]$ TeV.
 We show specific choices of $M$ as solid lines with varying $m_A$, in order to demonstrate the effect on $\kappa_3$ and $\kappa_4$.
 The gray (dark) shaded regions are the areas excluded by tree-level perturbative unitarity according to the analysis of Fig.~\ref{fig:unitarity}, and the shaded light blue region is defined in the same way as in  Fig.~\ref{fig:unitarity}. 
    } \label{fig:2hdm}
\end{figure*}

In order to present an example where Eq.~\eqref{eq:k3k4rel} can be realised, we consider the Two-Higgs Doublet Model (2HDM), where beyond tree level, the cubic and quartic self-couplings can receive significant contributions, as shown in~\citere{Bahl:2022jnx} (see also~\citere{Bahl:2023eau}). A review of the 2HDM can be found in \citere{Branco:2011iw}. We work in the alignment limit with the lightest scalar identified as the $125$~GeV Higgs boson (after Electroweak Symmetry Breaking (EWSB) and rotation to the Higgs basis) and perform a one-loop calculation\footnote{We use {\sc{FeynArts}}~\cite{Kublbeck:1990xc,Hahn:2000kx}, {\sc{FormCalc}}~\cite{Hahn:2001rv} and {\sc{LoopTools}}~\cite{Hahn:1998yk}, using the model of \citere{model}.} of the trilinear and quartic couplings 
employing the on-shell renormalisation scheme. As a motivated example we pick a benchmark point from~\citeres{Bahl:2022jnx,Bahl:2023fxi} which is compatible with the latest experimental results while also receiving sizeable trilinear-coupling corrections. We reproduce the one-loop result of 
\citeres{Bahl:2022jnx,Bahl:2023fxi}
and also show
the quartic coupling in Fig.~\ref{fig:loop}. As expected, the 
prediction for
the quartic coupling 
quickly rises to values even beyond what is allowed by tree-level perturbative unitarity in the $\kappa$-framework if the splitting between the mass of the CP-odd Higgs boson in the 2HDM, $m_A$, and the BSM Higgs scale $M = m_{12}/(c_\beta s_\beta)$ 
increases. In the displayed example the unitarity bound is violated if $m_A$ surpasses $\sim 1100$~GeV, and further two-loop contributions would tighten the bound on $m_A$.

In Fig.~\ref{fig:2hdm} the linear relation between $\kappa_3$ and $\kappa_4$ in the 2HDM is shown for variations of the scale $M$ 
and for masses $m_A \leq 1.5$~TeV. 
Varying the values of $m_A$ and $M$ shifts the relation between the self-couplings while maintaining 
a linear correlation between them. 
For $\kappa_3 =6$ the corresponding results for $\kappa_4$ vary between $\kappa_4 \approx 22$ and $\kappa_4 \approx 31$ for the displayed scenarios. Thus, the largest allowed values for $\kappa_3$ according to the present bounds are correlated in the 2HDM with very large shifts in $\kappa_4$. As indicated by the shaded light blue region in the plot these predictions for $\kappa_3$ and $\kappa_4$ are associated with a well-behaved power expansion within an EFT framework. 
While it would also be of interest to explore which models can induce an even larger deviation of $\kappa_4$ for relatively small values of $\kappa_3$, potentially resulting in regions that require a non-linear effective prescription (for instance the Electroweak Chiral Lagrangian), we leave such an investigation for future work.

\section{Triple Higgs production at the HL-LHC}\label{sec:hllhc}
\label{sec:hllhc}

The production of three Higgs scalars 
at the LHC and future colliders
is highly suppressed compared to single and double Higgs production, severely limiting the available final states that can be explored 
at the LHC.
In order to obtain the highest values for the product of cross section and branching ratios,
one needs to consider the dominant production mode through gluon fusion, but also the main decay channel to a $b$-quark pair. 
The latter is difficult in hadron collisions due to the sizeable multi-jet background from QCD processes. It can be 
problematic for typical cut-and-count analyses to sufficiently suppress the background while at the same time avoiding 
a large reduction of signal events in order to maximise significance. In this work we resort to Machine Learning (ML) techniques 
for 
appropriately 
selecting the signal region of 
the considered channels.

In order to identify which of the
decay channels of the on-shell Higgs bosons can be utilised for the analysis at the LHC, we 
start with an optimistic estimate of the number of events for the $6 b$, $4 b 2 \tau$, $2 b 4 \tau$ and $4 b 2 \gamma$ final states.\footnote{The different final states from triple-Higgs production have been 
discussed in Ref.~\cite{Papaefstathiou:2015paa}. While in principle channels with two $W$ bosons can also be of relevance, we choose not to explore them 
in view of the difficulty of the final states.}
Within the SM the involved branching ratios are given as
\begin{equation}
\begin{aligned}
& \BR(H \to b \bar{b}) = 0.584, \\
& \BR(H \to \tau^+ \tau^-) = 6.627 \times 10^{-2} \\
& \BR(H \to \gamma \gamma) = 2.26 \times 10^{-3} \;.
\end{aligned}
\end{equation}
We note that the $4 b 2 \gamma$ and $2 b 4 \tau$ final states only produce a few events at $3$/ab, even at relatively large coupling modifiers $\kappa_3 \gtrsim 4.5$, $\kappa_4 \gtrsim 30$ (taking into account K-factors of 1.7~\cite{deFlorian:2019app} and tagging efficiencies of all taus and all-but-one $b$-quarks). It is therefore unlikely that these channels 
will be statistically significant at the HL-LHC, even though they can be highly relevant for colliders utilising higher energies, as shown in~Refs.~\cite{Fuks:2015hna,Papaefstathiou:2015paa,Chen:2015gva,Bizon:2018syu}\footnote{For further studies of other channels at higher energies see also Refs.~\cite{Kilian:2017nio,Kilian:2018bhs}.
}. We therefore will not consider these channels further, and instead focus on the $6 b$ and $4 b 2 \tau$ channels. 

The background processes for the $6 b$ final state have been thoroughly discussed in Ref.~\cite{Papaefstathiou:2019ofh} (see also Ref.~\cite{Papaefstathiou:2020lyp}), and it is expected that the dominant contribution arises from multi-jet QCD $6b$ events. This is the only background that is taken into consideration for this final state in this work, and we neglect subdominant channels.

In the $4b2\tau$ channel\footnote{See~\citere{Fuks:2017zkg} for an analysis at FCC energies.}, the dominant backgrounds arise from the production of four $b$-quarks along with two $W$ bosons ($WWbbbb$) or one $Z$ boson ($Zbbbb$). The former includes the production of a top and bottom pair ($t t b b$) with subsequent decays $t \rightarrow W b$. The production of a top pair associated with a Higgs ($t t H$) or a $Z$ boson ($t t Z$) also yields noteworthy contributions. Here the $ttH$ channel is particularly problematic if a reconstructed resonance close to the $125$~GeV mass is required during an analysis to isolate the triple-Higgs signal. The final background included in our analysis is the four top production ($tttt$).

\subsection{Analysis}
\subsubsection{Event generation and pre-selection}\label{sec:gen}
We use {\sc{Madgraph5\_MC@NLO}}~\cite{Alwall:2014hca,Hirschi:2015iia} for event generation and modify the provided SM model file in the {\sc{UFO}}~\cite{Degrande:2011ua} format to introduce the modifications of the trilinear and quartic Higgs couplings $\kappa_3$ and $\kappa_4$, respectively.\footnote{We checked that our (loop-induced) leading order cross sections for the production of three 
(undecayed)
Higgs bosons
is in good agreement with Refs.~\cite{Plehn:2005nk,Binoth:2006ym}.
}  Signal events are generated for $p p \rightarrow h h h$ and are subsequently decayed on-shell with {\sc{Madspin}}~\cite{Artoisenet:2012st} 
in order to obtain the
cross section rates. Due to the 
complexity of the multi-particle final states we generate events with a minimum transverse momentum for the $b$-quarks of $p_T(b) > 28$~GeV 
and within the pseudorapidity region $\abs{\eta} < 2.5$, 
while we will later impose stricter cuts during the analysis. Additionally, since the signal consists of three on-shell Higgs bosons, we impose a cut on the invariant mass of the process of $\sqrt{\hat{s}} > 350$~GeV at generation level. 

While in principle one could explore different cuts in order to efficiently identify the signal region, the complexity of the final states would render this a cumbersome and difficult procedure, possibly requiring the use of complicated observables. Instead, we resort to Graph Neural Networks (GNNs) for 
an efficient discrimination between signal and background events. This requires an appropriate embedding of particle events to graphs. 
Before we address the ML aspects of the analysis it is appropriate to define pre-selection conditions required to be satisfied by each event that gets passed to the network. 

Showering and hadronisation is performed with {\sc{Pythia8}}~\cite{Sjostrand:2014zea} saving the resulting events as {\sc{HepMC}} files~\cite{Dobbs:2001ck}. {\sc{FastJet}}~\cite{Cacciari:2011ma,Cacciari:2005hq} is interfaced through {\sc{Rivet}}~\cite{Buckley:2019stt,Bierlich:2019rhm}, and jets are clustered using the anti-kT algorithm~\cite{Cacciari:2008gp} of radius $0.4$ and requiring a transverse momentum of $p_T(j) > 30$~GeV. We use {\sc{Rivet}} to calculate the events that will pass the pre-selection using a
$b$-tagging efficiency (independent of $p_T$)
of $0.8$. 
For the $6b$ ($4b2\tau$) channel, at least five (three) $b$-quarks are required, satisfying the conditions $p_T(b) > 30$~GeV and $\abs{\eta(b)} < 2.5$. For the $4b 2\tau$ channel two $\tau$ particles must also be identified in the central part of the detector, $\abs{\eta(\tau)} < 2.5$, with $p_T(\tau) > 10$~GeV. The $\tau$ particles are identified with  the {\sc{TauFinder}} class 
of {\sc{Rivet}},
and at least one 
$\tau$ particle must decay hadronically.\footnote{We take inspiration from the $bb\tau\tau$ analysis of Ref.~\cite{ATLAS:2022xzm}.} We apply an efficiency of $0.8$ for both leptonic and hadronic taus.\footnote{It should be noted that such efficiencies have already shown to be achievable, see e.g.\ Ref.~\cite{CMS:2022prd}.} The invariant mass of the sum of the four-momenta of the above final states should 
exceed $350$~GeV, otherwise the event is vetoed. 
Finally, we form combinations of $b$-quark pairs, and at least one pair is required to have an invariant mass close to the mass of the Higgs boson, $m_{b \bar{b}} \in \left[110, 140\right]$~(in GeV). In the case of the $4b 2 \tau$ channel the event passes the pre-selection also if the invariant mass criterion is satisfied by the invariant mass of the di-tau state, $m_{\tau \tau}$.

\subsubsection{Graph Embedding and Neural Network Architecture}
\label{sec:nn}
GNNs, stemming from the idea that certain types of data can be 
efficiently represented as graphs, have been increasingly utilised in particle physics. Various works have indicated their applicability for BSM-relevant tasks such as event classification~\cite{Blance:2020ktp,Abdughani:2020xfo}, jet-tagging~\cite{Qu:2019gqs,Dreyer:2020brq}, particle reconstruction~\cite{Pata:2021oez}, identifying anomalies in data arising from BSM interactions~\cite{Atkinson:2022uzb,Atkinson:2021nlt} and obtaining constraints on parameters in SMEFT or the $\kappa$-framework~\cite{Atkinson:2021jnj,Anisha:2022ctm}.\footnote{For a detailed
list of references in particle physics using GNNs, see Ref.~\cite{Feickert:2021ajf}.} The latter is what we aim to achieve by performing a fit within the $\kappa_3$--$\kappa_4$ plane after the efficient selection of a signal region from the GNN. Similar architectures using graphs have also been recently utilised by experiments at the LHC, see e.g.\ Ref.~\cite{ATLAS:2023ajo}.

The generated events need to be embedded in graphs before they are passed to the neural network. We explore two different paths\footnote{All considered graphs are bidirectional.}: 
\begin{enumerate}
	\item{\label{itm:fc} \textit{Fully Connected (FC):} Add nodes for all the considered final states 
    (i.e.~$b$~quarks and $\tau$ leptons, denoted as $b_i$ and $\tau_i$ according to their $p_T$ values) and edges connecting all the nodes. We use the transverse momentum, pseudorapidity, azimuthal angle, energy, mass and PDG identification number as node features, $\left[p_T, \eta, \phi, E, m, \text{PDGID}\right]$, while no edge features are introduced. A node is also added for the missing momentum of the event.
}
	\item{\label{itm:ph} \textit{Reconstructed Nodes (RN):} Add fully connected nodes for $b$~quarks (and $\tau$ leptons for the $4 b 2\tau$ final state) as before, but additionally add nodes $H_i$ for reconstructed pairs of particles $i$, $j$ that are (relatively well) compatible with the Higgs-boson mass, $m_{ij} = 125 \pm 25$~GeV. This is achieved by forming combinations between all the $b$-quarks and (if applicable) the $\tau$-pair. The $H_i$ nodes correspond to the four-momentum and mass of the reconstructed pair, ordered according to which is closest to the Higgs-boson mass of $125$ GeV. All the nodes have $\left[p_T, \eta, \phi, E, m, \text{PDGID}\right]$ as associated features, where the PDGID for $H_i$ is zero. }
\end{enumerate}
Such physics-inspired approaches according to the expected chain of the event have been shown to improve results in semi-leptonic top decays~\cite{Atkinson:2021jnj} and are actively being explored~\cite{Ehrke:2023cpn}. 

GNNs operate by calculating messages using node features (and edge features if these exist) and iteratively updating the node features for each message passing layer. We rely on the EdgeConv~\cite{edgeconv} operation for message passing, where the message of node $i$ at the $l$-th message passing layer is calculated with
\begin{equation}
	\label{eq:message}
	 m_{ij}^{(l)} = \text{\sc{ReLU}} \left(\Theta(\vec{x}_j^{\,(l)} - \vec{x}_i^{\,(l)}) + \Phi(\vec{x}_i^{\,(l)})\right)\;,
\end{equation} 
where $\Theta$ and $\Phi$ indicate linear layers. The node features for $l=0$ are the kinematical quantities we have defined as inputs, and the updated node features are obtained from the messages by averaging over the neighbouring nodes, 
\begin{equation}
	\label{eq:nodereadout}
	\vec{x}_i^{(l+1)} = \frac{1}{\abs{{\cal{N}}(i)}} \sum_{j \in {\cal{N}}(i)} m_{ij}^{(l)}\;.
\end{equation}
The final node features after all EdgeConv operations are aggregated to a single vector using a `mean' graph readout operation. In principle, it is possible to additionally include further (non-graph related) layers at this stage. The final network score is obtained with a linear layer with the SoftMax activation~\cite{softmax} that reduces the resulting features to a two-dimensional vector, with each entry representing the probability that the event was signal or background. The amount of EdgeConv and the following linear layers need to be optimised to achieve high performance while at the same time avoiding overfitting. After experimenting with different setups, we settled on using two EdgeConv layers with hidden features of size $96$ before the output layer for both channels.

\subsubsection{Training and comparison of graph embeddings}\label{sec:embeddingscomp}

The data is split into subsamples of $\sim 56\%$, $19\%$ and $25\%$ for training, validation and testing, respectively\footnote{In total we include 22000 events for each class for the $5b$ case and 35000 events for the $3b2\tau$ case.}, and we minimise the cross-entropy loss function in order to train the network using the {\sc{Adam}} optimiser~\cite{DBLP:journals/corr/KingmaB14}. The learning rate is one of the hyperparameters requiring tuning, and for our case the value of $0.001$ ($0.01$) performs best
for $3b2\tau$ ($5b$). If for three epochs in a row the loss has not decreased, then the learning rate is reduced
by a factor of $0.1$. In principle the training can run for up to $200$ epochs, although we impose early stopping conditions if the loss has not improved for ten consecutive epochs. A batch size of $128$ is used for every update of the loss function.

The GNN for the $5b$ analysis is trained on two classes (signal and background).  The situation is more involved for the $3b 2 \tau$ 
case where the analysis benefits significantly from a multi-class classification which allows identifying different thresholds for the different background scores. In particular we choose to train on the $W W b b b b$, $Z b b b b$ and $t \bar{t} (H \rightarrow \tau^+ \tau^-)$ contributions. The signal events used for training are always for the $(\kappa_3, \kappa_4) = (1,1)$ point (using different values does not significantly alter the performance of the network).

\begin{figure*}[t!]
\subfigure{\includegraphics[width=8.3cm]{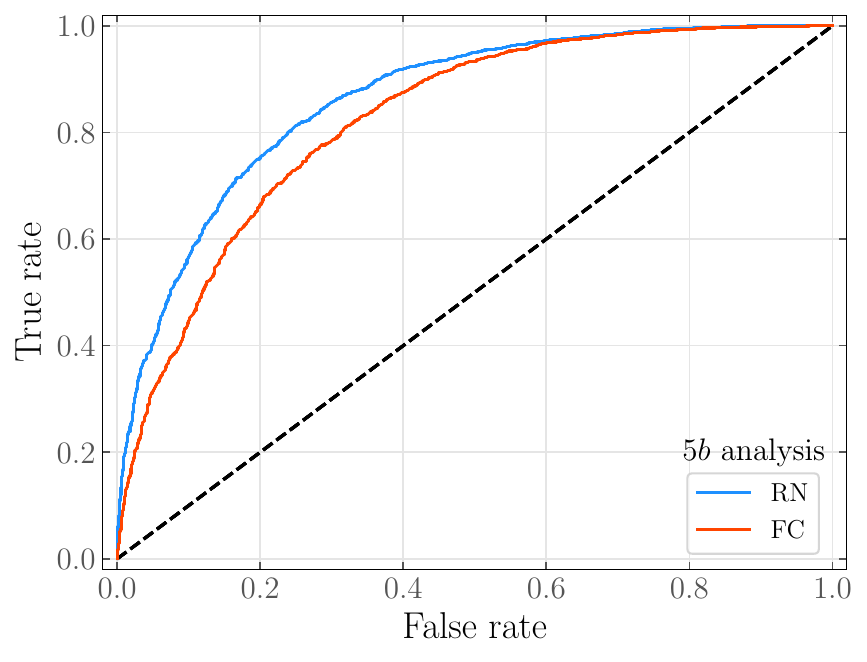}}
\hfill
	\subfigure{\includegraphics[width=8.3cm]{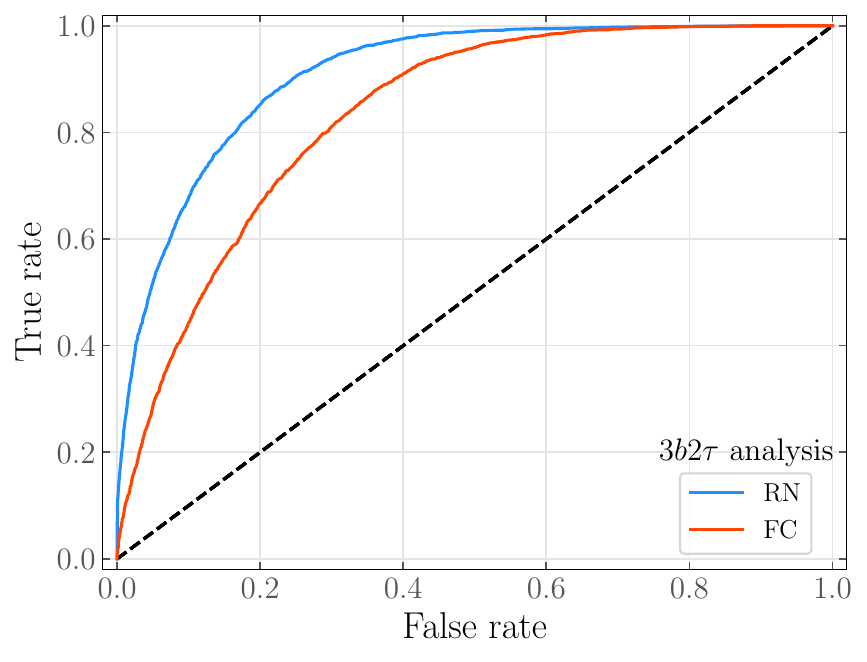}}
	\caption{ROC curves for the $5b$ and $3b2\tau$ analyses are 
    displayed on the left and right, respectively, showing the performance of the two embedding cases. For the $3b2\tau$ case we binarise in a one-vs-rest 
    scheme, grouping together the background classes. The areas under the `RN' and `FC' curves for the $5b$ case are 0.862 and 0.823, respectively. For the $3b2\tau$ analysis the `RN' area is 0.909 and the `FC' area is 0.833.
    \label{fig:roc}}
\end{figure*}

We use the EdgeConv implementation from the {\sc{Deep Graph Library}}~\cite{wang2020deep} with {\sc{PyTorch}}~\cite{paszke2019pytorch} as backend. The graph embedding relies on {\sc{PyLHE}}~\cite{pylhe} to extract events from the Les Houches Events (LHE) files~\cite{Alwall:2006yp}. In order to compare the different graph embeddings, we use functionality from {\sc{scikit-learn}}~\cite{scikit-learn} to calculate the true and false positive rates at different thresholds\footnote{As we perform multi-class classification for the $3b2\tau$ analysis, we binarise the output of the network for the purpose of this comparison.}, and we show the corresponding Receiver Operating Characteristic (ROC) curves for both channels in Fig.~\ref{fig:roc}. 

The ROC curves and the distributions allow one to conclude that the RN embedding utilising the reconstructed Higgs 
boson mass
can lead to significant improvements. This is not unexpected as additional information (available at detector level) is passed to the network to aid classification. While in principle a sufficiently deep
neural network with fully-connected graphs could also eventually learn to map the input features of the $b$-jets (and taus) to the masses of the reconstructed Higgs bosons, including the information in the graph embedding allows easier optimisation and quick convergence with a shallow network. We therefore utilise only the RN embedding for performing the final signal region selection. 

\subsubsection{HL-LHC Results}
For simplicity we will use the signal efficiency of the network for the $(\kappa_3, \kappa_4) = (1,1)$ point and assume that it will be mostly the same irrespective of the coupling modifier values as our analysis is largely dependent on the cross section rates. Ideally one could train and optimise a network on each point, or alternatively train on event samples from topologies that depend differently on $\kappa_3$ and $\kappa_4$.\footnote{As a simple test we also trained on a sample that includes two signal classes for $(\kappa_3, \kappa_4) = (1,0)$ and $(0,1)$ for the $3b2\tau$ analysis (effectively separating diagrams with $\kappa_3$ and $\kappa_4$ insertions) and tested this setup on different points. While there is an improvement in efficiency for certain points (up to $~10\%$), for most cases of $(\kappa_3, \kappa_4)$ the efficiency is closer to the simpler setup with one signal class. We chose to utilise the latter, since it also enables an easier interpretation of our results in Sec.~\ref{sec:interpparton}.} 
For the $5b$ analysis we optimise the signal selection 
to reduce the false positive rate to $\sim 0.6\%$.
In the $3 b 2 \tau$ channel we require the following conditions to be satisfied on the background network scores:
\begin{itemize}
	\item $\text{P}[WWbbbb] < 3\%$\;,
	\item $\text{P}[Zbbbb]  < 10\%$\;,
	\item $\text{P}[t\bar{t}(H \rightarrow \tau^+ \tau^-)] < 30\%$\;. 
\end{itemize}
It should be noted that even though the network was trained only on a subset of possible background contributions, it still performs well, as discussed below,
and manages to remove background contributions from other sources as well. We calculate the efficiencies for each contribution and show the reduction of cross sections in Tab.~\ref{tab:bkgs}. Our results for both channels include a K-factor for the signal of $1.7$~\cite{deFlorian:2019app} and a conservative estimate of the higher-order contributions to the background 
processes in terms of a K-factor of $2$. 

\begin{table}
\begin{tabular}{llll}
\toprule
	{} & $\sigma (\text{gen.}) (\text{fb})\;\;$ & $\sigma (\text{sel.}) (\text{fb})\;\;$  & $\sigma (\text{NN}) (\text{fb})$ \\
\midrule
	$t t (h\rightarrow \tau \tau)$    &   $3.3$ &  $0.14$  &  $0.011$ \\
	$w w b b b b$					  &   $27$  &   $4.0$  &  $7.1\times10^{-3}$ \\
	$t t (h \rightarrow b b)$         &   $3.0$ &  $0.78$  &  $3.3\times10^{-3}$ \\
	$z b b b b$                       &   $3.8$ &  $0.40$  &  $2.9\times10^{-4}$ \\
	$t t (z \rightarrow b b)$         &  $0.67$ &  $0.13$  &  $2.7\times10^{-4}$ \\
    $t t t t$						  &  $0.33$ &  $0.080$ &  $1.8\times10^{-4}$ \\
	$t t (z \rightarrow \tau \tau)$   &   $4.1$ &  $0.073$ &  $9.0\times10^{-5}$ \\
\bottomrule
\end{tabular}
	\caption{Background contributions included in the $3b2\tau$ analysis and reduction of the generated cross sections (labelled as ``gen.'') after pre-selection cuts (``sel.'') and GNN selection (``NN''). $W$-bosons arising from tops are allowed to decay hadronically and $c$-jets can be mis-tagged as $b$-jets with a probability of $0.2$. \label{tab:bkgs}}
\end{table}

\begin{figure*}[t!]
\subfigure{\includegraphics[width=8.3cm]{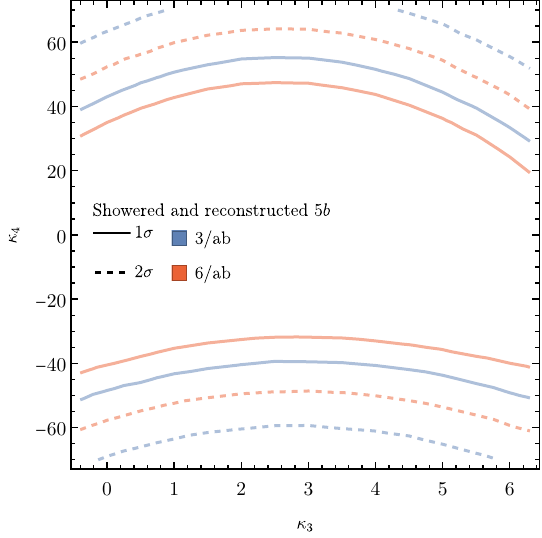}}
\hfill
\subfigure{\includegraphics[width=8.3cm]{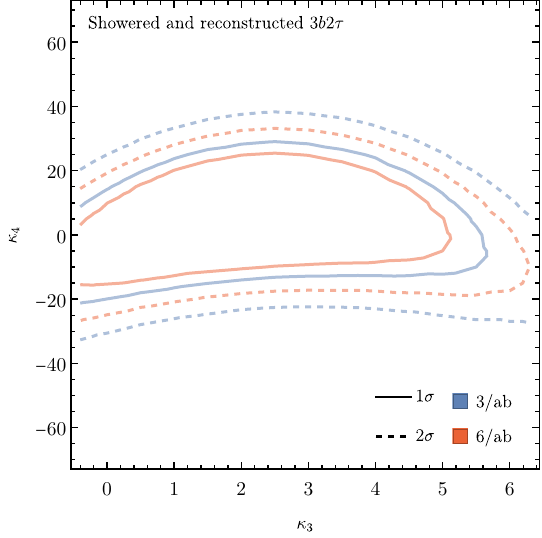}}
	\caption{Projected contours indicating the $1\sigma$ and $2\sigma$ bounds in the $\kappa_3$--$\kappa_4$ plane from the $5b$ (left) and the $3 b 2 \tau$ (right) analysis, including effects from showering, hadronisation and reconstruction. \label{fig:contoursreco}}
\end{figure*}

We define the significance for our analysis 
according to~\cite{Cowan:2010js}
\begin{equation}
    Z = \sqrt{2 \bigg( (S + B) \ln{(1+ \frac{S}{B})} - S\bigg)} \;,
\end{equation}
where $S$ and $B$ denote the signal and background events,
respectively. 
This allows us to obtain $1\sigma$ and $2\sigma$ bounds within the $\kappa_3$--$\kappa_4$ plane (which roughly correspond to $68\%$ and $95\%$ CL, respectively), as shown in Fig.~\ref{fig:contoursreco} for the $5b$ and $3 b 2 \tau$ analyses. We assume an integrated luminosity at the HL-LHC of $3$/ab and a combined ATLAS and CMS luminosity of $6$/ab. 

Overall, we observe that the $3 b 2 \tau$ analysis is more sensitive than the $5 b$ analysis, and the latter will additionally suffer from further subdominant electroweak contributions to the background that have not been included.\footnote{We have tested our trained network on a sample of $Zb\bar{b}b\bar{b}$ events and note that the resulting cross section of this background in the signal region is $\sim 3\%$ of the QCD background. We 
do not include triple boson backgrounds (e.g.~$ZZZ$, $HZZ$ and $HZZ$) which have comparable cross sections to the SM $HHH$ process~\cite{Hirschi:2015iia}. Using the conservative assumption that such backgrounds will have an efficiency similar to the signal implies ${\cal{O}}(1)$ events which would be negligible compared to our included backgrounds.}

\begin{figure*}[t!]
	\subfigure{\includegraphics[width=0.45\textwidth]{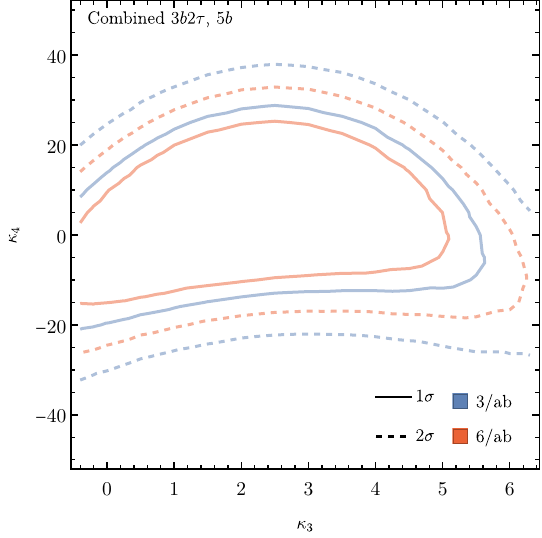}}
	\hfill
	\subfigure{\includegraphics[width=0.45\textwidth]{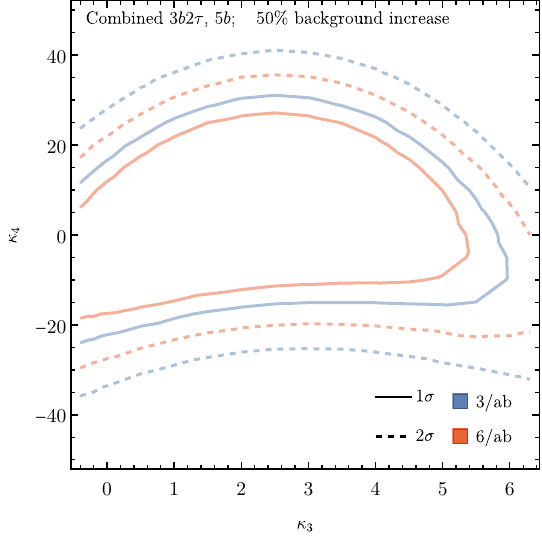}}
	\caption{The left plot shows the projected contours 
    indicating the $1\sigma$ and $2\sigma$ bounds in the $\kappa_3$--$\kappa_4$ plane obtained from a combination of the $5b$ and $3 b 2 \tau$ channels under the assumption that there are no correlations. 
    The right plot shows the corresponding result where the backgrounds for both channels are increased by 50\%.
    \label{fig:contourcomb}}
\end{figure*}

However both channels should be utilised in combinations to maximise the significance. Assuming for simplicity zero correlations between the channels, we combine the significances as $Z_\text{comb} = \sqrt{Z_{5 b}^2 + Z_{3 b 2 \tau}^2}$, giving rise to the contours shown in Fig.~\ref{fig:contourcomb} (left).
While the projected bounds of about $\pm 20$ times the predicted value for the quartic Higgs self-coupling in the SM may appear to be quite weak, in view of our discussion above we emphasise that such bounds go much beyond the existing theoretical bounds. Furthermore, deviations of this size in $\kappa_4$ are well compatible with the existing experimental bounds on $\kappa_3$ according to the correlations between $\kappa_3$ and $\kappa_4$ that are present in the BSM scenarios analysed above. Regarding the sensitivity to $\kappa_3$ from triple Higgs boson production at the HL-LHC, Fig.~\ref{fig:contourcomb} shows that the expected sensitivity in this channel at the HL-LHC is weaker than the present experimental limits that have been derived from di-Higgs production. Combining this independent set of experimental information on $\kappa_3$ with the experimental results from di-Higgs production may nevertheless turn out to be useful.
While our analysis may be optimistic in some respects (e.g.~we neglect fake taus), on the other hand we note that further developments of the triggers, tagging and reconstruction algorithms of final states 
could result in
higher efficiencies than 
the values that we have adopted in our analysis,
enhancing the significance. The ability to discriminate between jet flavours is highly important for $HHH$ studies (as well as $HH$ studies) and could also allow experiments to study fully hadronic final states where $H$ decays to $W$ bosons. 
On the other hand, we note that
even in the case that the backgrounds are increased by $50\%$, the resulting constraints on $\kappa_3$ and $\kappa_4$ degrade only slightly,
as shown in Fig.~\ref{fig:contourcomb} (right).

\subsection{Interpretability of NN scores}
\label{sec:interpparton}

Understandably, NN techniques are often viewed as ``black boxes'', due to their inability to indicate the input features that 
are most important for determining 
their predicted scores. 
In order to address this shortcoming,
various approaches have been explored in the recent years 
with the goal to yield
interpretability, allow efficient debugging of the network, better understand the mapping between input and output, and ultimately allow the identification of ways to improve it. These methods gained traction in particle physics in the recent years to 
obtain a better insight for various different tasks such as jet- and top-tagging and detector triggers~\cite{deOliveira:2015xxd,Chang:2017kvc,Agarwal:2020fpt,Chakraborty:2019imr,Andreassen:2019txo,Mahesh:2021iph,Khot:2022aky}.

\begin{figure*}[!t]
	\includegraphics[width=0.95\textwidth]{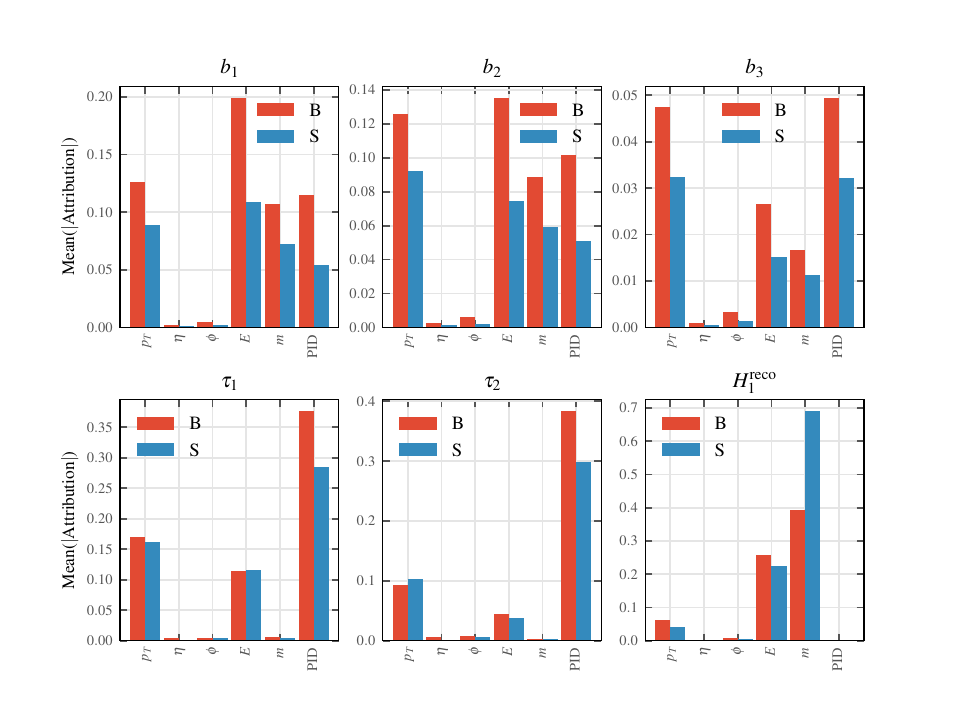}
	\caption{Attributes for the features of the $b$~jets, $\tau$ leptons and the reconstructed invariant Higgs-boson mass that is closest to the $125$~GeV resonance. The height of the attribution value indicates to what extent 
    the network is using the particular feature in order to discriminate between signal and background. In the figure we denote collectively all the background classes as `B' and the signal as `S'.} 
    \label{fig:attributions}
\end{figure*}

There are various techniques 
for gaining interpretability in ML, but in general they can be separated into two categories: intrinsically interpretable models that are specifically designed to increase transparency providing intuition and post-hoc explanation methods that were developed to enhance our understanding of generic ML 
models. The latter is what applies to the case of this work. However, many post-hoc techniques lack certain properties that are beneficial to maintain; for example one could directly use the product of the gradients computed during backpropagation and the input in order to attribute the most relevant features~\cite{JMLR:v11:baehrens10a,2013arXiv1312.6034S}. As the gradients of the network hold information regarding variations of the inputs, 
it should be possible to use them
to quantify the dependence of the score on features. It is known, though, that gradient methods can yield the same attribution for an input and a baseline that differ 
from each other
and have different outputs (for an example see Ref.~\cite{integratedgradients}), due to the gradient becoming flat (this is often the case as NNs are trained until the loss is saturated). 

Shapley~\cite{shapley} values (originating from Game Theory), are formulated based on certain axioms to distribute the attributions amongst the participating variables in a ML approach and have been applied for obtaining interpretations~\cite{shap} 
(for an application in particle physics, see Refs.~\cite{Grojean:2020ech,Alasfar:2022vqw,Grojean:2022mef}).  Their attractiveness stems from the fact that they follow axiomatic principles unlike earlier methods (e.g.~DeepLift~\cite{deeplift} or Layer-wise relevance propagation~\cite{lrp}). However, their evaluation is often computationally expensive  and requires multiple calls of the neural network. 

Integrated Gradients (IGs) is an alternative approach,  designed in Ref.~\cite{integratedgradients} using axiomatic considerations, which requires significantly fewer calls to the network function. The trade-off is the requirement 
that the ML technique must be differentiable, which is the case for NNs optimised through gradient descent approaches, and 
the application of IGs
also requires access to the gradient of the model\footnote{Often techniques such as Shapley values are called ``black-box'' approaches as they have no access to anything other than the output of the ML approach, while IGs and similar techniques are refered to as ``white-box'' approaches.}. Let a generic classification NN  denoted as $F: \mathbb{R}^{n} \rightarrow  \left[0, 1\right]$ for input features $x \in \mathbb{R}^n$ and $x^\prime \in \mathbb{R}^n$ denote an appropriate baseline (e.g.~a zero vector). Integrating over all the gradients of $F$ in a straight path from $x^\prime$ to $x$ defines IGs as
\begin{equation}
	\label{eq:ig}
	{\cal{I}}_i(x) = (x_i - x_i^\prime) \int_{0}^{1} d\alpha \frac{\partial F(x^\prime + \alpha (x - x^\prime))}{\partial x_i}\;.
\end{equation}

\begin{figure*}[t!]
\subfigure{\includegraphics[width=8.3cm]{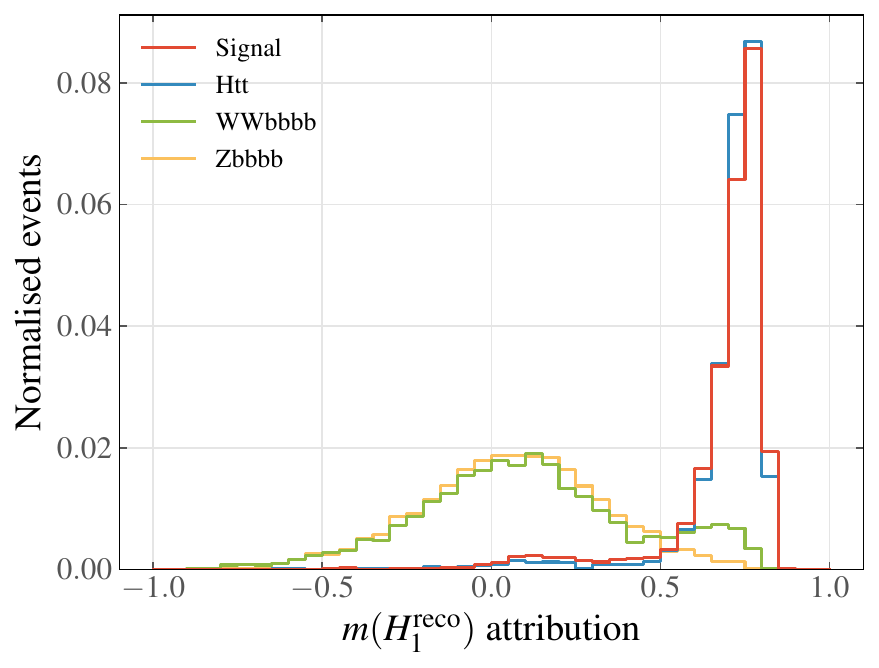}}
\hfill
\subfigure{\includegraphics[width=8.3cm]{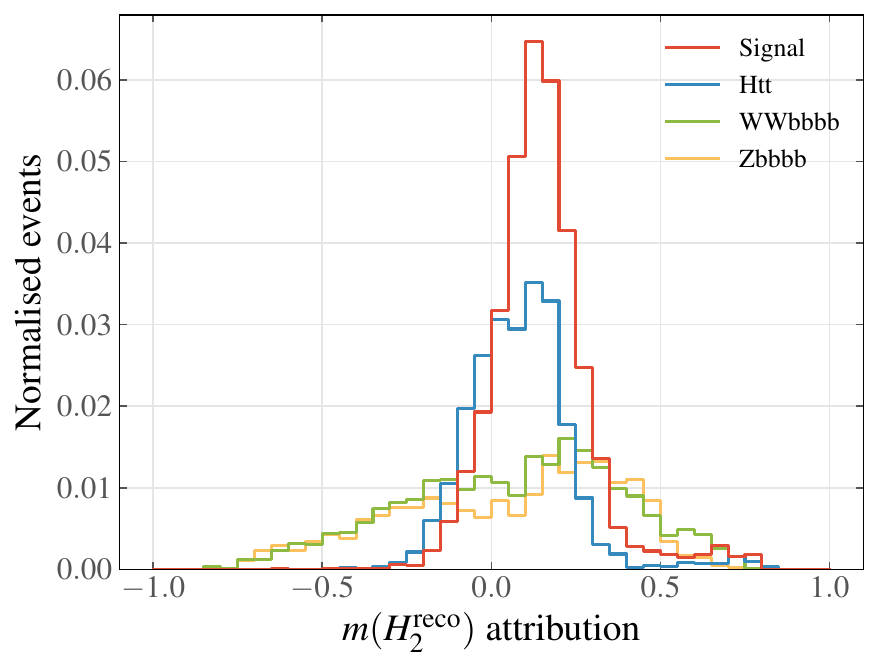}}
    \caption[test]{Histograms showing the attribution for different events against the value of the reconstructed mass for the (true) signal and backgrounds. The plot on the left (right) shows the reconstructed mass of the Higgs boson that is closest (second-closest) to the $125$~GeV resonance. A positive attribution close to 1 indicates events with a high output score (i.e.\ identified as signal), while lower values of the
    attribution imply a low output score.
    \label{fig:Hvsattr}}
\end{figure*}

We thus utilise IGs, implemented in the {\sc{Captum}}~\cite{captum} library, in order to obtain attributions for our predictions and identify the most relevant inputs for our processes.

The attributions obtained from IGs allow us to interpret the results of the network in terms of the input parameters for each node, as shown in Fig.~\ref{fig:attributions}, although 
some care is necessary
when interpreting such results. Quite intuitively the transverse momenta and the energy of the b-jets are relevant parameters 
that receive high attributions. This is expected since 
restricting to higher values of $p_T$ can help in the discrimination between signal and background
(this was also the reason for applying 
a pre-selection momentum cut). Angular momenta are not so helpful for discrimination; this is not unexpected as we are dealing with scalars. The network additionally utilises the PID of the tau 
leptons more than the 
identification of the
$b$~quarks; this is likely 
due to the fact that the di-tau state is correlated with the highly discriminative reconstructed invariant mass of the Higgs boson.  
We clearly see that the introduction of the reconstructed masses 
significantly boosts
the performance of the network, being the most important observable for the signal events. We note that as a reconstructed particle, the Higgs node 
has been assigned a PID of zero which as required by the `dummy' axiom\footnote{The `dummy' axiom states that a variable that is not contributing to the output of the network should have no attribution, ensuring that the attribution is insensitive to irrelevant inputs. It is a standard axiom imposed by interpretation methods (see e.g.~\citeres{integratedgradients,shapley}).} has no attribution and thus zero contribution to the network results.

Taking a closer look at the reconstructed masses and their attributions, we see in Fig.~\ref{fig:Hvsattr} that the node with a reconstructed mass 
that is closest to
125~GeV receives a sizeable attribution.\footnote{We note that the reconstructed quantity that is closer to the actual mass of the observed Higgs boson can be the di-tau state, which is less affected 
by showering the events than for the case where the Higgs boson decays
to $b$~quarks
and can yield more events closer to the actual mass of the Higgs boson.} 
The attributions from the mass of the $H_1^\text{reco}$ node 
indicate that due to the similarity of the $t\bar{t}(H\rightarrow\tau^+ \tau^-)$ background with the signal, the network is unable to clearly discriminate the two classes based on this feature alone. The inclusion of the mass of the second reconstructed Higgs boson, however, helps the network as indicated by the higher attributions assigned to the signal events as compared to the other sources of backgrounds. This implies that the inclusion of reconstructed observables can enhance the performance of GNNs in certain analyses, as also expected from the discussion in Sec.~\ref{sec:embeddingscomp}.

We stress that while the IG attributions provide an 
indication of the most important variables, our approach does not
yield detailed information on
how the specific correlations between the input features can impact the network score. While in many cases this would be desired, this is beyond the scope of our work where we use IGs as a method to verify that the introduction of the reconstructed Higgs-boson mass is indeed the most relevant variable. We leave explorations of alternative techniques (also specific to GNNs) pinpointing to important connections between input features and nodes for future work.

In our work we utilised interpretation methods mostly to ensure that the GNN works as expected and in order to identify potential issues during the implementation of the network. However, the usefulness of such techniques extends well beyond this. For example, in the case of limited computing resources one could check which features are irrelevant and remove them from the analysis before scaling the network up. Indeed, in our 
analysis we checked
that if we remove the seemingly unused angular information, we obtain similar results as before (resulting in no visible changes in Fig.~\ref{fig:contoursreco} for $3b2\tau$). 
Additionally for analyses with multiple final states the most practical observable 
that can be exploited
is not always straightforward to identify. Interpretation techniques could therefore be used as a first step to 
identifying the most relevant
observables before optimising the analysis to enhance its significance.

\section{Reach assessment for lepton colliders and comparison with the HL-LHC}
\label{sec:llcol}

For 
comparison with the prospects of the HL-LHC, we finally consider the expected upper limits on $\kappa_3$ and $\kappa_4$ from possible future lepton colliders.\footnote{This topic has previously been explored in Refs.~\cite{Maltoni:2018ttu,Gonzalez-Lopez:2020lpd,Chiesa:2020awd}.} We consider an inclusive analysis of $\ell \ell \rightarrow H H H + \text{other}$ which includes both the associated $Z H H H$ production and the production through WBF. In principle one could consider dedicated analyses for each channel, optimising the selection of final states; however, we choose to perform an inclusive analysis to avoid further assumptions on the identification of other states which could vary depending on the collider concept and the detector. We will consider the decay 
$H \rightarrow b \bar{b}$
of the Higgs boson, which yields the largest possible cross section for the signal, and assume throughout that the $b$-tagging efficiencies will be $0.8$. Our analysis relies solely on identifying $5 b$ jets in the clean environment provided by lepton collisions. We apply an additional $\sim 0.83$ efficiency which arises from requiring the $p_T$ of the $b$~jets to be larger than $30$~GeV. We note that in practice the 
results for an electron or muon collider would be similar, 
i.e.\ the obtained contours for the limits in the $\kappa_3$--$\kappa_4$ plane for a given collider c.m.\ energy and integrated luminosity would not be expected to significantly differ for the two collider types.
Therefore we will refer to generic lepton colliders in the following, although we use the centre-of-mass energies of $1$ and $3$ TeV envisaged for the ILC and CLIC, as well as $10$ TeV collisions that could be realised at a muon collider. We scan over different values of $\kappa_3$ and $\kappa_4$ for the aforementioned energies and subsequently apply the relevant tagging efficiencies. 

\begin{figure}[t]
	\includegraphics[width=0.45\textwidth]{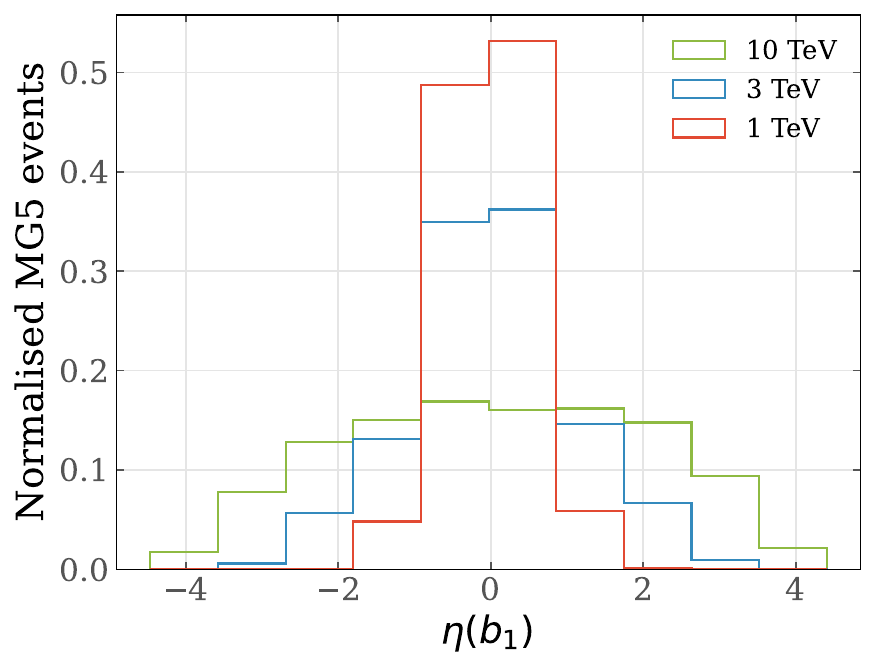}
	\caption{Pseudorapidity distribution for the leading $b$~quark for 
    different collision energies. \label{fig:leprap}}
\end{figure}

An important limitation of high-energy lepton collisions in this case, however, 
arises from the region where the detectors can tag $b$~jets. While for energies $\sim 1$~TeV the $b$~quarks are in the central part of the detector, the situation is significantly different for $10$~TeV collisions, as shown in Fig.~\ref{fig:leprap}. It is thus necessary to explore possibilities for extending the tagging capabilities of future detectors to even $\abs{\eta} \sim 4$ in order to avoid a significant loss of events.

\begin{figure*}[t!]
\subfigure{\includegraphics[width=8.3cm]{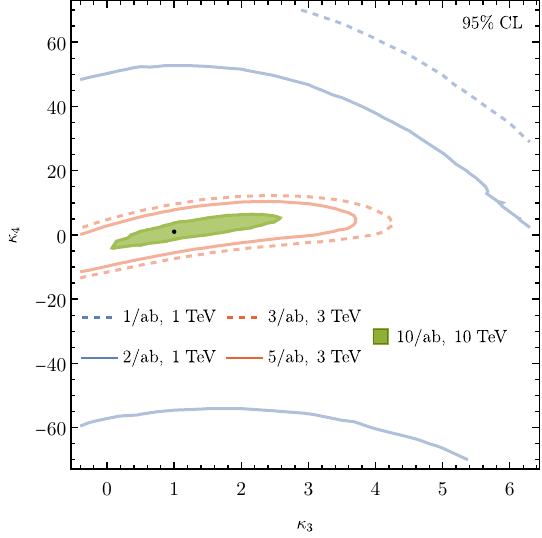}}
\hfill
\subfigure{\includegraphics[width=8.3cm]{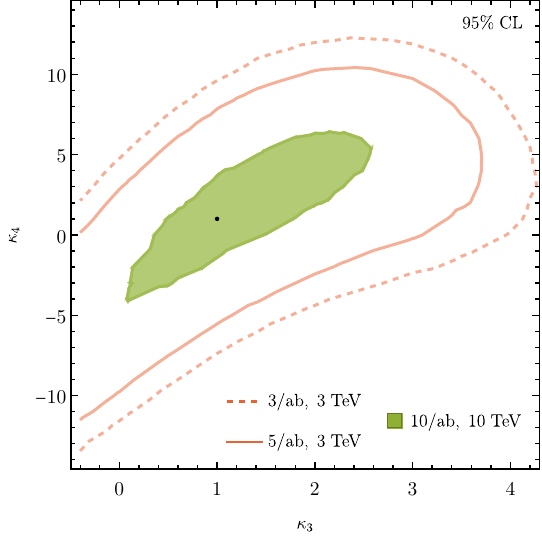}}
	\caption{On the left, the projected 95\% CL contours for lepton colliders at different energies and integrated luminosities are shown, mainly focusing on the energies of ILC, CLIC and a possible muon collider. The SM value is shown as a black dot. The plot on the right shows a zoomed-in version. \label{fig:lepcoll}}
\end{figure*}
\begin{figure*}[!t]
	\includegraphics[width=0.65\textwidth]{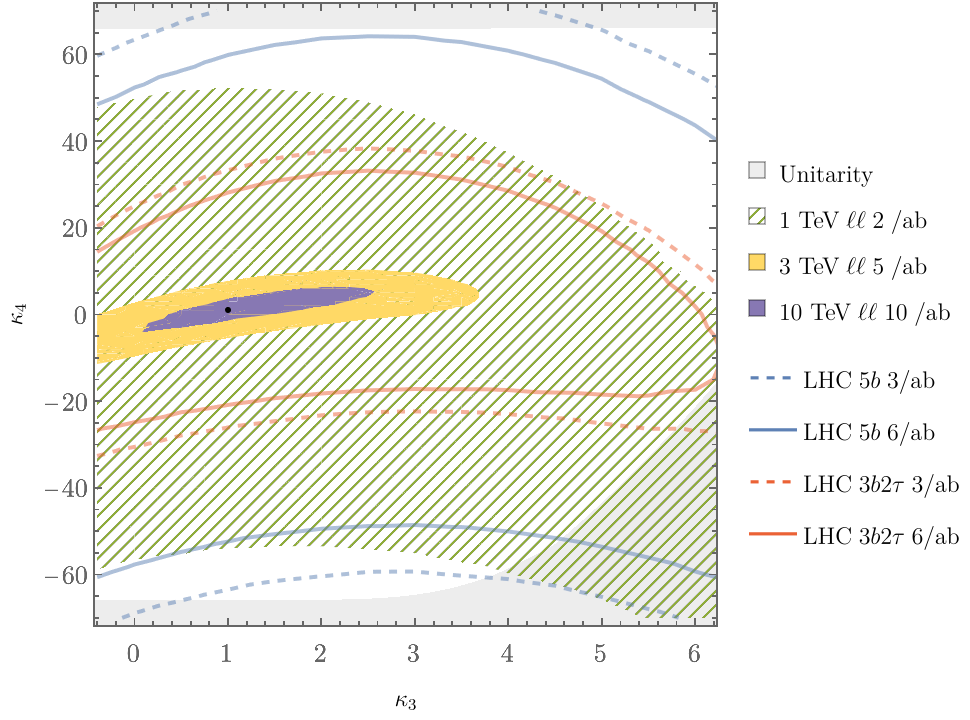}
	\caption{Comparison of the projected 95\% CL contours for the $5 b$ and $3 b 2 \tau$ analyses at the HL-LHC with the projected 95\% CL sensitivities at lepton colliders with different energies (indicated by the different coloured regions). The shaded gray area indicates the region that is excluded by the bound from tree-level perturbative unitarity.\label{fig:all}}
\end{figure*}

The leptonic collisions deliver clean signals avoiding the large background contamination from QCD that is present at hadron colliders. We checked the leading order QCD background of the signature $(5 b + \text{other})$ and found that the cross sections of these background processes are small. Assuming that the selection of the signal region will enforce $p_T(b) > 30$~GeV, the requirement that one di-bottom pair should be compatible with the mass of the observed Higgs boson and a cut ensuring that the total invariant mass of the final state particles produced in the process is at least $350$~GeV would result in no remaining background events (even with more relaxed cuts, the number of events is negligible when taking $b$-tagging efficiencies into account). However, similar to Refs.~\cite{Maltoni:2018ttu,Gonzalez-Lopez:2020lpd,Chiesa:2020awd} we do not take into account electroweak backgrounds which could be dominant and deserve a dedicated study. We turn to a Poissonian analysis as described in Ref.~\cite{ParticleDataGroup:2022pth}, where $n$ corresponds to the number of events expected from the SM, i.e.\ for $(\kappa_3, \kappa_4) = (1,1)$. 
Upper limits on the mean value of the Poisson distribution $\mu$ are then calculated with
\begin{equation}
	\label{eq:upperlim}
	\mu_\text{up} = \frac{1}{2} F^{-1}_{\chi^2} \bigg[ 2 (n+1) ; \text{CL} \bigg]\;,
\end{equation}
where $F^{-1}_{\chi^2}$ denotes the inverse of the cumulative distribution of the $\chi^2$ distribution, and CL is the confidence level (e.g.~95\%). 

The resulting bounds at $95\%$~CL are shown in Fig~\ref{fig:lepcoll} for different centre-of-mass energies and integrated luminosities. The plots show that the large luminosities expected to be utilised by colliders at $3$~TeV and $10$~TeV (as envisaged for CLIC and muon-colliders, see~Refs.~\cite{CLIC:2018fvx,Accettura:2023ked}) provide significant constraining power via the triple Higgs production process for $\kappa_4$ and $\kappa_3$. 

The lepton collider projections are compared with our results for the HL-LHC in Fig.~\ref{fig:all}. We find that the HL-LHC sensitivity for $\kappa_4$ is competitive with the one achievable at a $1$~TeV lepton collider such as the ILC. In particular the comparison shows that for negative $\kappa_4$ the HL-LHC is expected to have a better sensitivity than a $1$~TeV lepton collider.

As discussed above further developments in ML could increase both the tagging and selection efficiencies beyond our assumptions, and additional channels will provide additional information.

\section{Conclusions}
\label{sec:conc}

Our investigation of the prospects at the HL-LHC shows that even though triple-Higgs production is limited by low rates at the LHC, its exploration
provides interesting information even if it does not receive additional contributions from 
new scalar resonances. Bounds can be placed on $\kappa_4$ significantly beyond the theoretical constraints from perturbative unitarity. 

While as expected the bounds on $\kappa_3$ will be much weaker than the ones from double-Higgs production, they should be useful for improving the sensitivity through combinations. Additionally, if deviations from the SM are found, the correlation between the Higgs self-couplings can shed light on the possible scenarios of physics beyond the SM.

If an excess in the triple Higgs production process is observed,
the correlation with the result for double-Higgs production will be immensely informative. On the one hand, if no 
deviation from the SM value
is identified in $\kappa_3$ from other channels, 
an indication for a large 
deviation in $\kappa_4$  would likely imply the presence of 
non-linear effects
that cannot be described consistently within an effective field theory approach via the expansion in terms of a heavy scale. On the other hand, a deviation in both coupling modifiers could indicate a correlation between $\kappa_3$ and $\kappa_4$ 
that can be confronted with predictions of specific 
models such as the 2HDM and of effective field theories. 

The physics gain that can be achieved via
the statistically limited channel of triple Higgs production 
at the HL-LHC  
crucially depends on an efficient signal--background discrimination. 
For this purpose we have employed in our analysis
the use of GNNs. It is already evident from current experimental searches that such ML techniques will be the centerpiece of future studies. However, it is especially important in particle physics to be able to identify the relevant kinematical features that contribute to the identification of the signal. An unintuitive behaviour (e.g.\ a high-attribute quantity that is already known to be irrelevant)
could indicate a possible issue in the learning framework. Alternatively, potentially interesting quantities could be identified that could 
provide discriminative power even in simpler analyses that do not use ML algorithms. 
We have explored 
interpretability
within GNNs using IGs which satisfy necessary axioms. 
We have shown that, as expected, the invariant mass of bottom and tau pairs 
is the most important feature in the data that is utilised for discrimination. We expect that such techniques  
will play an important role
not only for the development of analyses for BSM searches but also for further applications in particle physics.

Our comparison of the prospects at the HL-LHC with future lepton colliders shows that the sensitivity to $\kappa_4$ at the HL-LHC should be competitive with a $1$~TeV lepton collider such as ILC. While the sensitivities of lepton colliders at $3$ and $10$~TeV (e.g.\ CLIC or a possible muon-collider) are expected to be considerably higher, these results will presumably become available only on a longer time scale, such as the one for a future higher-energetic hadron collider. Thus, it can be expected that the HL-LHC will be able to establish the first bounds on $\kappa_4$ beyond theoretical considerations.

\bigskip
\noindent{\bf{Acknowledgements}} ---
We thank Henning Bahl, Akanksha Bhardwaj, Johannes Braathen, Philipp Gadow, Ulrich Haisch, Greg Landsberg, J\"urgen Reuter and Tania Robens for useful comments and discussions.
This work is supported by the Deutsche Forschungsgemeinschaft (DFG, German Research Foundation) under Germany’s Excellence Strategy EXC2121 “Quantum Universe” - 390833306 and has been partially funded by the Deutsche Forschungsgemeinschaft (DFG, German Research Foundation) - 491245950.

\bigskip
\noindent{\bf{Note added:}} ---
Shortly after our paper, Ref.~\cite{Papaefstathiou:2023uum} appeared on the {arXiv} which studies triple Higgs production in the $6b$ final state at HL-LHC and FCC, including coupling modifications beyond the Higgs self-couplings.

\bibliography{paper.bbl} 

\end{document}